\documentclass[a4paper]{jpconf}
\usepackage{graphicx}
\begin{document}
\title{Global p-mode oscillations throughout the complete solar cycle 23 and the beginning of cycle 24}

\author{D Salabert$^{1,2}$, R A Garc\'ia$^3$, P L Pall\'e$^{1,2}$ and A Jim\'enez$^{1,2}$}

\address{$^1$ Instituto de Astrof\'isica de Canarias, E-38200 La Laguna, Tenerife, Spain}
\address{$^2$ Departamento de Astrof\'isica, Universidad de La Laguna, E-38206 La Laguna, Tenerife, Spain}
\address{$^3$ Laboratoire AIM, CEA/DSM-CNRS, Universit\'e Paris 7 Diderot, IRFU/Sap, Centre de Saclay, F-91191 Gif-sur-Yvette, France}

\ead{salabert@iac.es, rgarcia@cea.fr, plp@iac.es, ajm@iac.es}

\begin{abstract}
The parameters of the p-mode oscillations vary with solar activity. Such temporal variations provide insights for the study of the structural and dynamical changes occurring in the Sun's interior throughout the solar cycle. We present here a complete picture of the temporal variations of the global p-mode parameters (excitation, damping, frequency, peak asymmetry, and rotational splitting) over the entire solar cycle 23 and the beginning of cycle 24 as observed by the space-based, Sun-as-a-star helioseismic GOLF and VIRGO instruments onboard SoHO. 
\end{abstract}

\section{Introduction}
The solar p-mode parameters have been demonstrated to vary with time and to be closely correlated with the solar activity proxies (Chaplin et al. [1]; Salabert et al. [2, 3]; Komm et al. [4]). These temporal variations provide insights to infer the interior of the Sun and its structural and dynamical changes throughout the solar cycle. However, clear differences between p-mode frequencies and solar activity during the unusually minimum of cycle~23 have been reported (Broomhall et al. [5]; Salabert et al. [6]). The origin of the p-mode variability is thus far from being properly understood and a better comprehension of its relationship with solar (and stellar, Garc\'ia et al. [7]) activity cycles will help us in our understanding of the dynamo processes.

\section{Observations and analysis}
\label{sec:anal}
We analyzed observations collected by the space-based, Sun-as-a-star instruments Global Oscillations at Low Frequency (GOLF, Gabriel et al. [8]) and Variability of Solar Irradiance and Gravity Oscillations (VIRGO, Fr{\"o}hlich et al. [9]) onboard the {\it Solar and Heliospheric Observatory} (SoHO) spacecraft. GOLF measures the Doppler velocity in the D1 and D2 sodium lines (Garc\'ia et al. [10]).
VIRGO is composed of three Sun photometers (SPM) at 402~nm (blue), 500~nm (green), and 862~nm (red). A total of 5202 and 5154 days of GOLF and VIRGO observations respectively covering more than 14 years between 1996 and 2010 were analyzed, with respective duty cycles of 95.4\% and 95.2\%.
These datasets were split into contiguous 365-day subseries, with a four-time overlap. The power spectrum of each subseries was fitted to extract the mode parameters (Salabert et al. [11]) using a standard likelihood maximization function. Subseries with duty cycles less than 90\% (basically around the SoHO vacation, 1998--1999) were removed. Each mode component was parameterized using an asymmetric Lorentzian profile, including the $l=4$ and 5 modes when visible. The temporal variations of the mode parameters were defined as the difference between reference values (taken as the average over 1996--1997) and the parameters of the corresponding modes observed at different dates. Their weighted averages over the central part of the 5-minute oscillation power ($\simeq $2200 -- 3400~$\mu$Hz) were then calculated.  Mean values of daily measurements of the 10.7-cm radio flux were used as a proxy of the solar surface activity. Linear regressions were performed between the temporal variations of the mode parameters and the radio flux using independent points only. The color code in the following figures corresponds to the VIRGO blue, green, and red channels respectively.

\begin{table}
\caption{\label{tab:paramchanges} Changes (\%) from maximum-to-minimum of the solar cycle in mode amplitudes $\langle\Delta h\rangle$, linewidths $\langle\Delta \gamma\rangle$, acoustic power $\langle\Delta p\rangle$, and energy supply rate $\langle\Delta \dot{e}\rangle$ measured from VIRGO data.}
\begin{center}
\begin{tabular}{llll}
\br
$\langle\Delta h\rangle$ & $\langle\Delta \gamma\rangle$  & $\langle\Delta p\rangle$  & $\langle\Delta \dot{e}\rangle$ \\
\mr
$-37.9 \pm 2.1$ & $19.9 \pm 1.6$ & $-18.1 \pm 1.0$ & $1.6 \pm 1.7$\\
\br
\end{tabular}
\end{center}
\end{table}

\begin{table}
\caption{\label{tab:changeasym} Changes (\%) from maximum-to-minimum of the solar cycle of the peak asymmetry $\langle\Delta b\rangle$ and their correlations with activity (using independent points only). }
\begin{center}
\begin{tabular}{lll}
\br
Instrument & $\langle\Delta b\rangle$  & Correlation\\
\mr
GOLF blue period & $-0.45 \pm 0.17$ & $-0.57$\\
GOLF red period   & $-0.30 \pm 0.33$ & $-0.74$\\
VIRGO blue            & $-0.69 \pm 0.22$ & $-0.55$\\
VIRGO green         & $-0.52 \pm 0.22$ & $-0.50$\\
VIRGO red              & $-0.44 \pm 0.21$ & $-0.38$\\
\br
\end{tabular}
\end{center}
\end{table}

\section{Results}
\subsection{Mode excitation and damping}
\begin{figure*}
\begin{center} 
\includegraphics[scale=0.31]{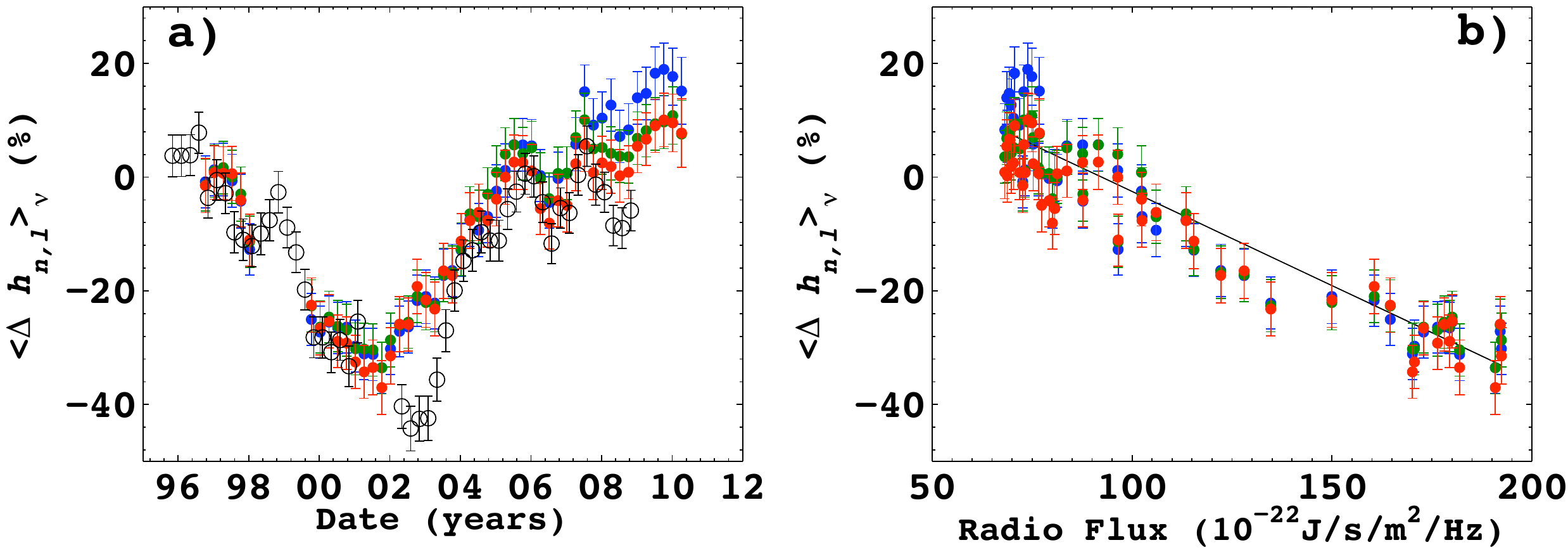}  \includegraphics[scale=0.31]{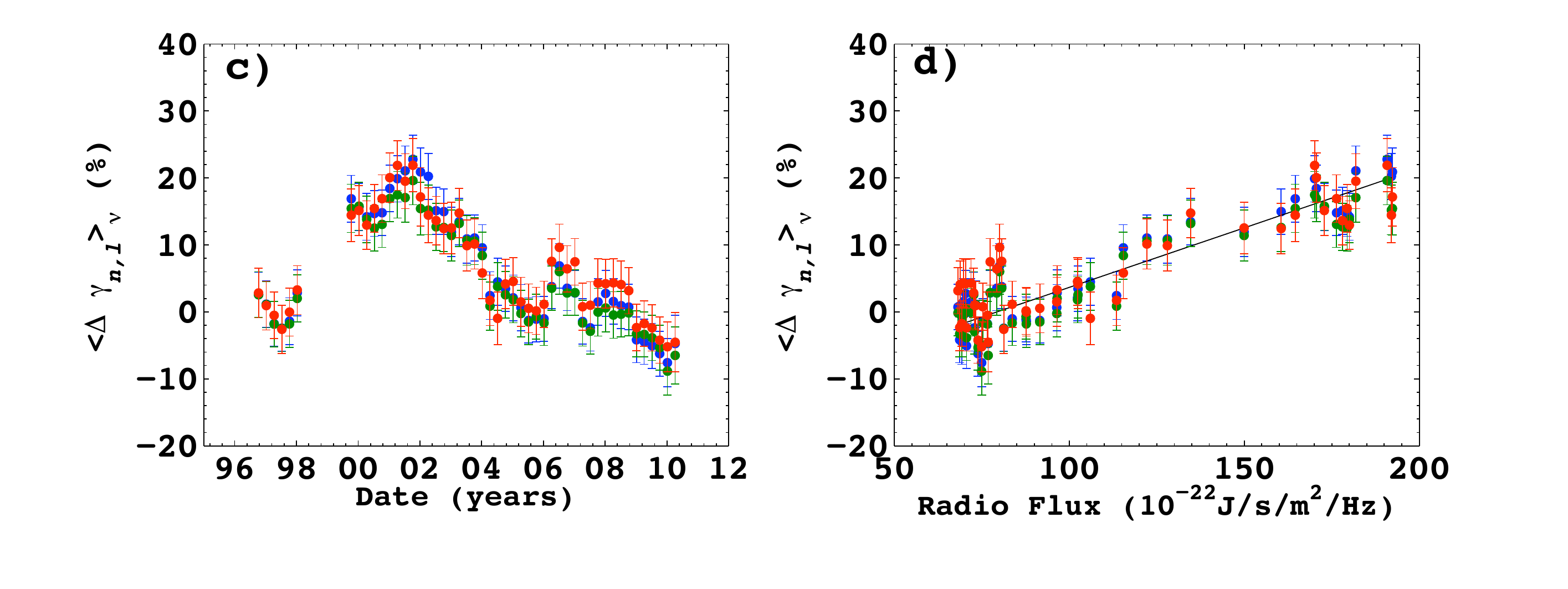} 
\end{center} 
\caption{\label{fig:virgo_amp} Temporal variations (\%) in mode amplitudes $\langle\Delta h\rangle$ (a, b) and linewidths $\langle\Delta \gamma\rangle$ (c, d) measured from VIRGO data as a function of time and  radio flux. The solid lines in panels b and d represent the best fits from weighted linear regressions.} 
\end{figure*}

The temporal variations of the mode excitation and damping parameters from VIRGO were averaged over the $l = 0$, 1 and 2 modes. The changes in mode amplitudes $\langle\Delta h\rangle$ and linewidths $\langle\Delta \gamma\rangle$ are shown on Fig.~\ref{fig:virgo_amp}. Note that due to absolute calibration problems and the changes of the observing wings, the GOLF amplitudes and linewidths are not exploitable for the moment. A proper calibration is currently underway. In order to compare with VIRGO, we performed, as described in Sec.~\ref{sec:anal}, a preliminary analysis of the Global Oscillation Network Group (GONG, Harvey et al. [12]) $l = 0$ data ({\tt gong.nso.edu/data/}). Similar fluctuations from the year 2006 in mode amplitudes (open circles on Fig.~\ref{fig:virgo_amp}) and linewidths are observed as in VIRGO. The changes from maximum-to-minimum of the solar cycle are given in Table~\ref{tab:paramchanges} and are consistent with previous work (e.g., Jim\'enez-Reyes et al. [13]; Salabert and Jim\'enez-Reyes [14]). 

\begin{figure*}
\begin{center} 
\includegraphics[width=1.38in]{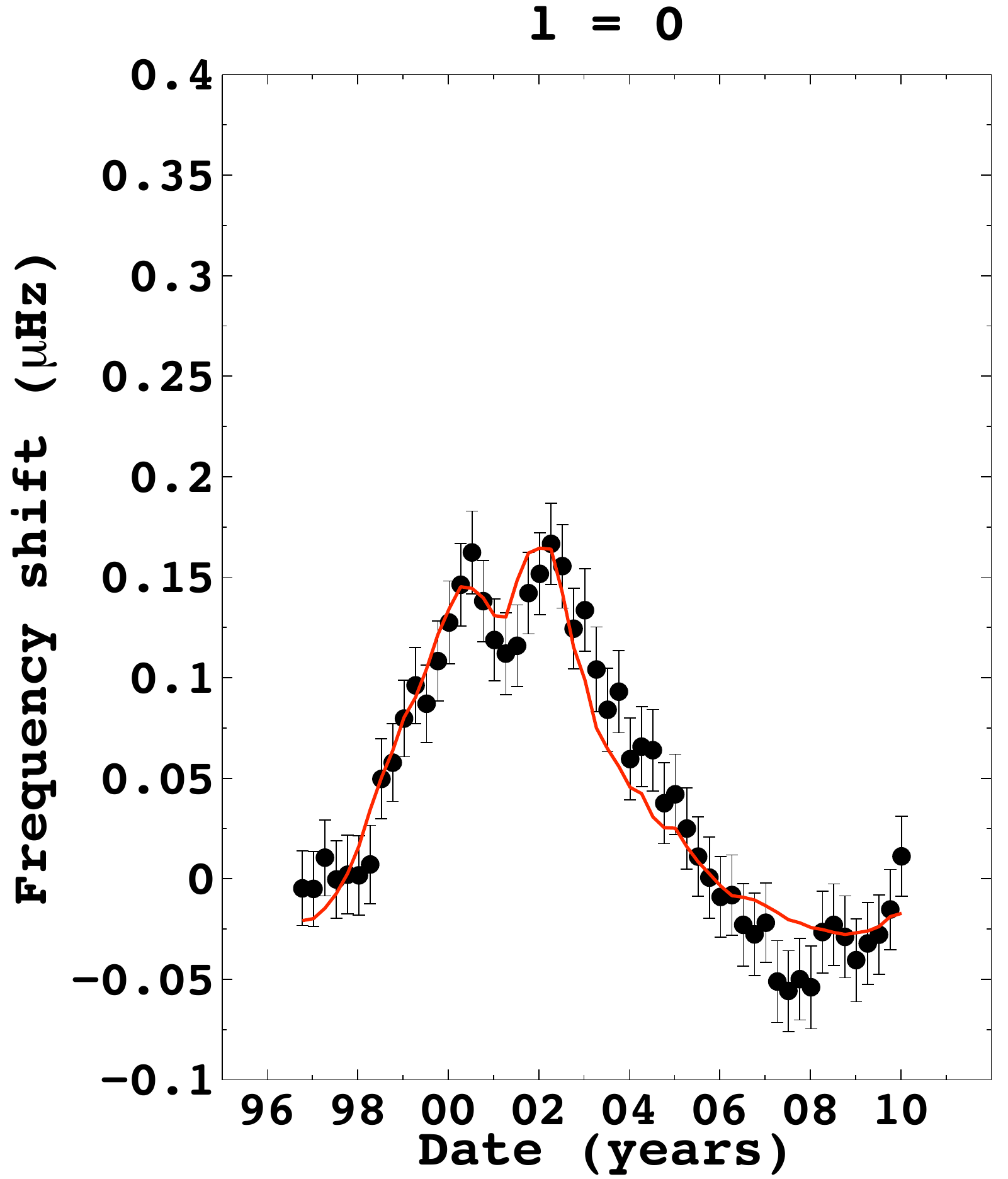}  \includegraphics[width=1.37in]{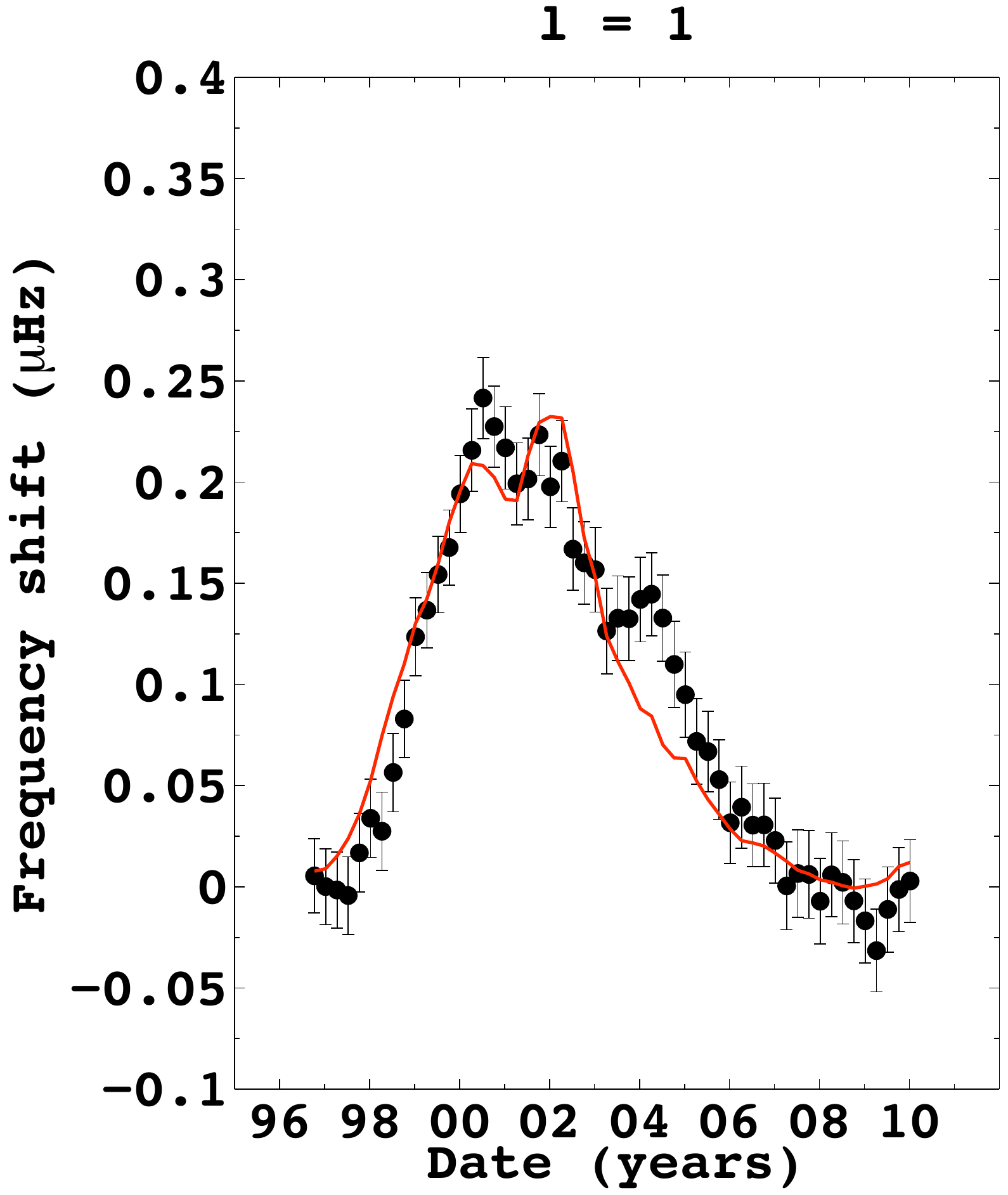}  \includegraphics[width=1.37in]{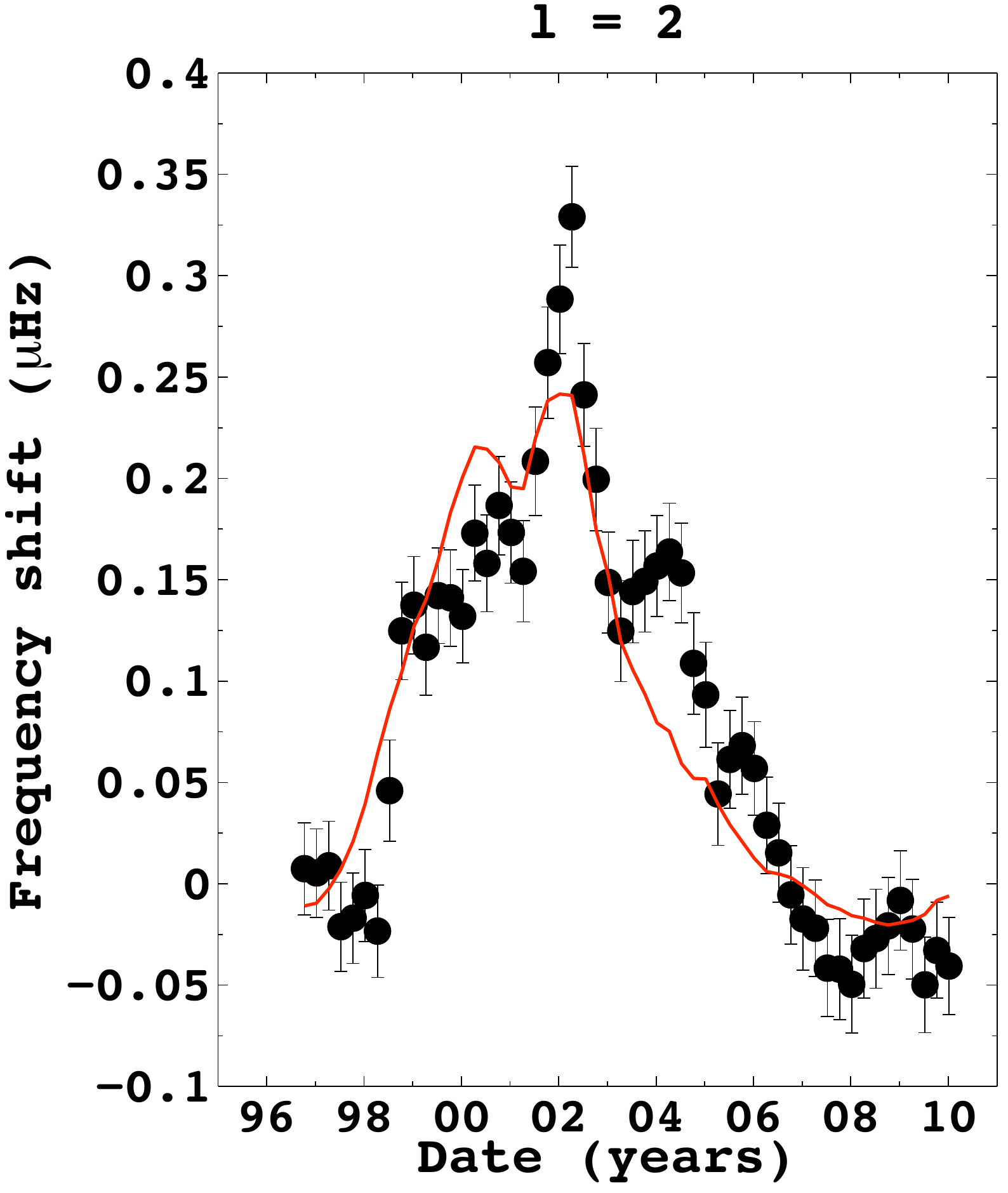} \\
\includegraphics[width=1.35in]{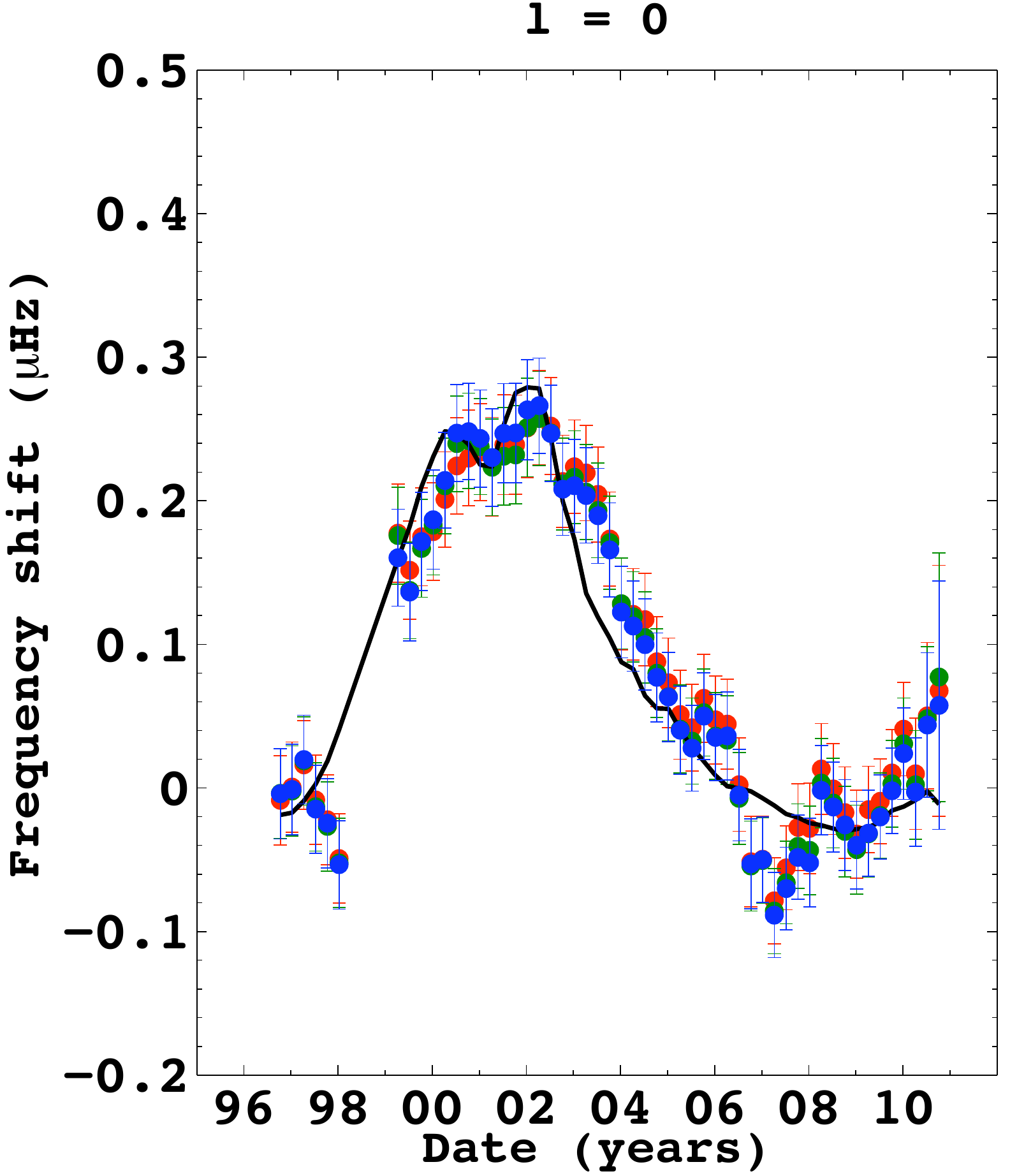}  \includegraphics[width=1.35in]{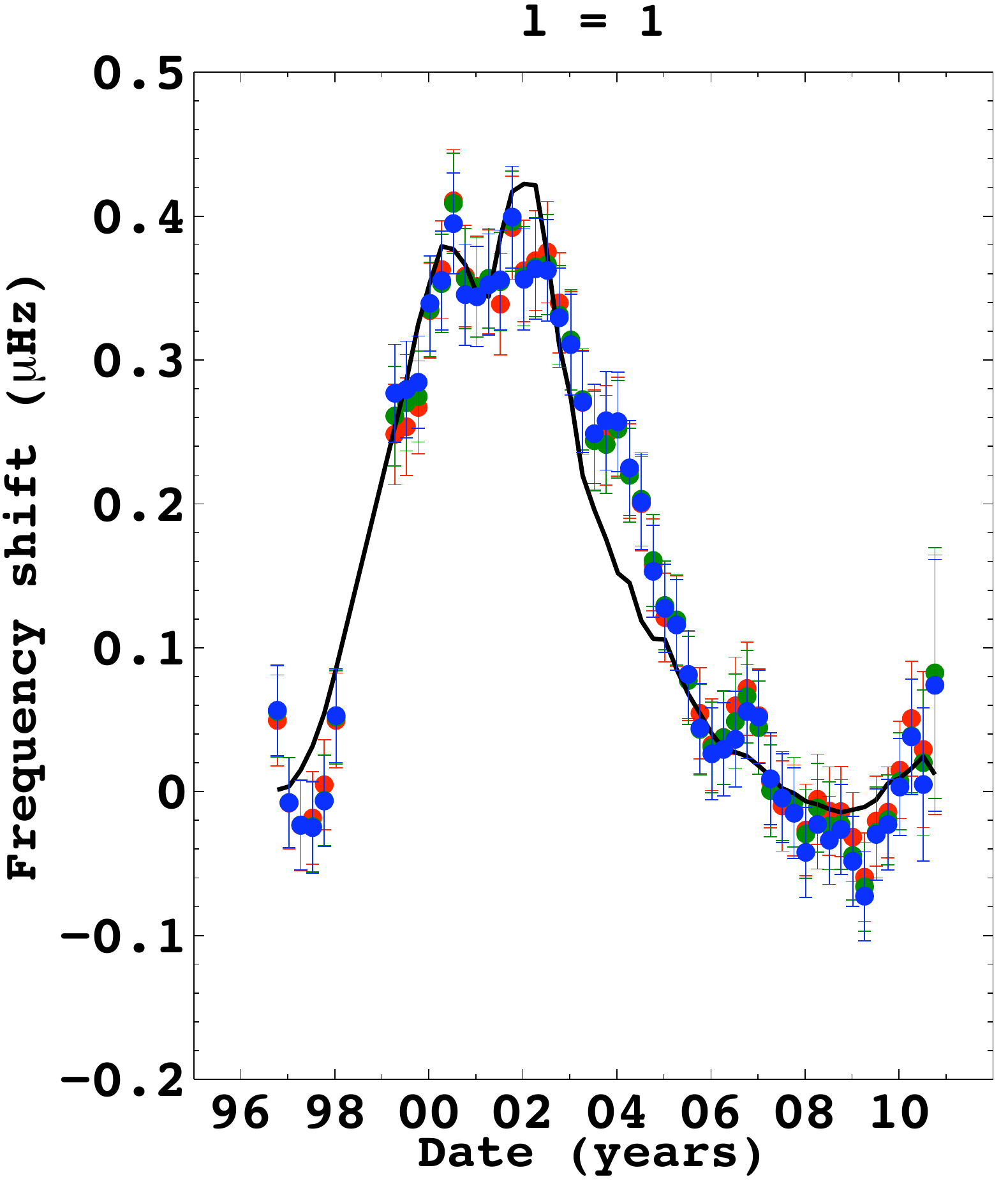}  \includegraphics[width=1.33in]{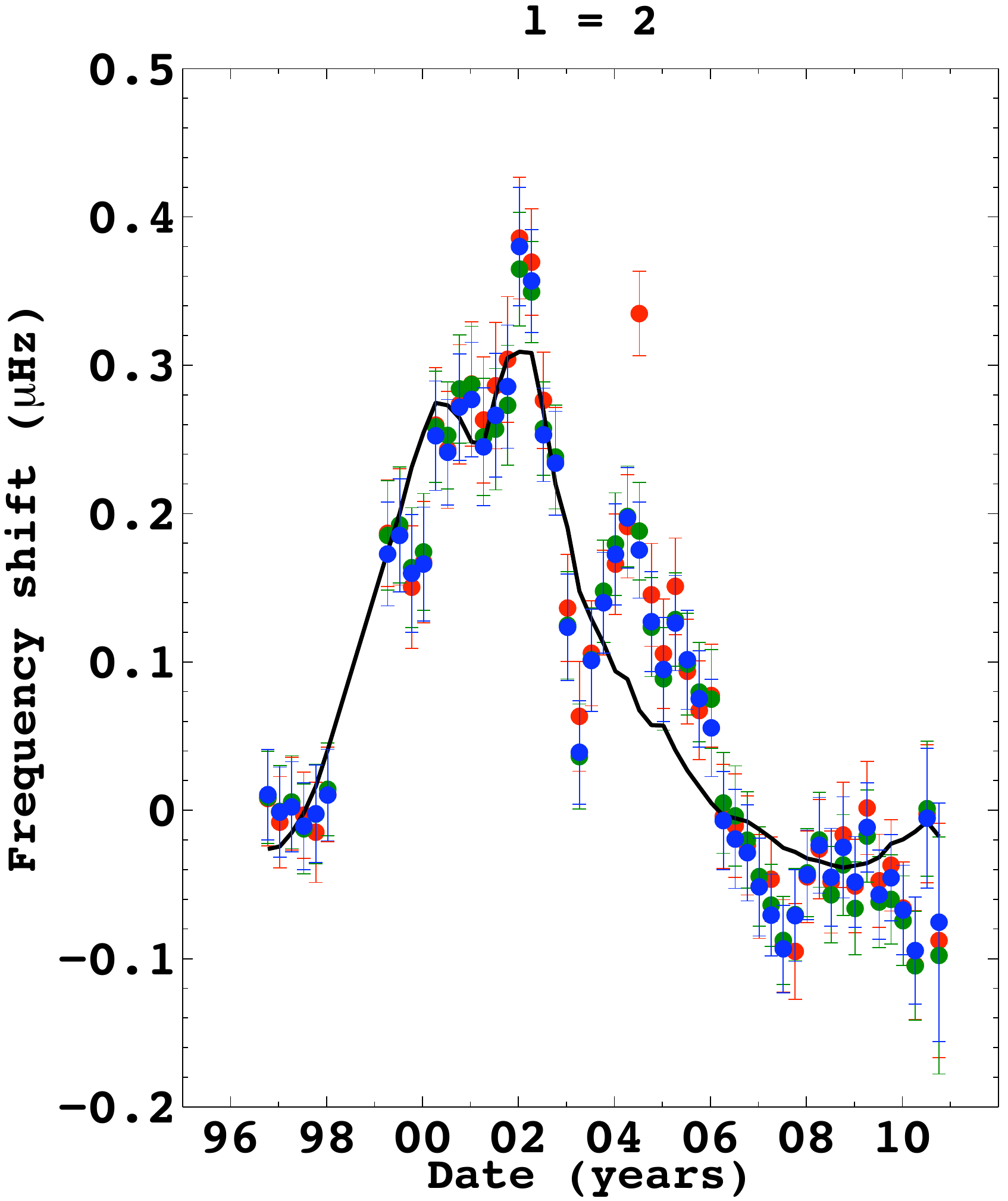} 

\end{center} 
\caption{\label{fig:golf_freq} Temporal variations ($\mu$Hz) of the $l=0$, 1, and 2 p-mode frequencies measured from GOLF (top) and VIRGO (bottom) data. The solid lines correspond to the scaled radio flux.} 
\end{figure*}

\begin{figure}
\begin{center} 
\includegraphics[scale=0.269]{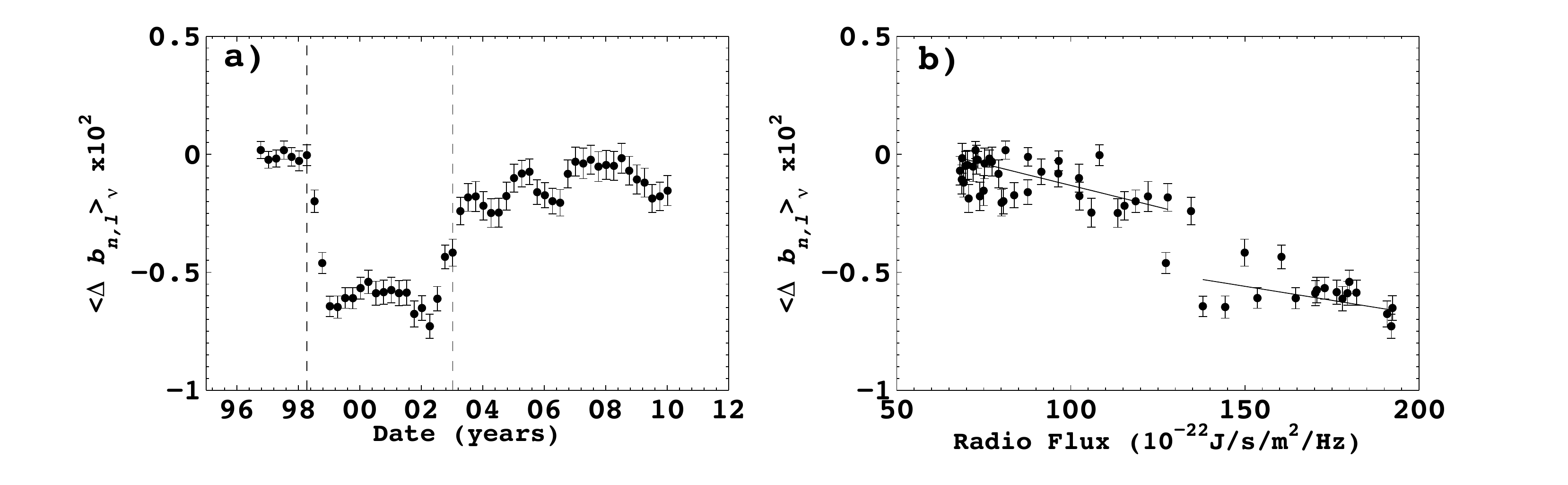}  \includegraphics[scale=0.261]{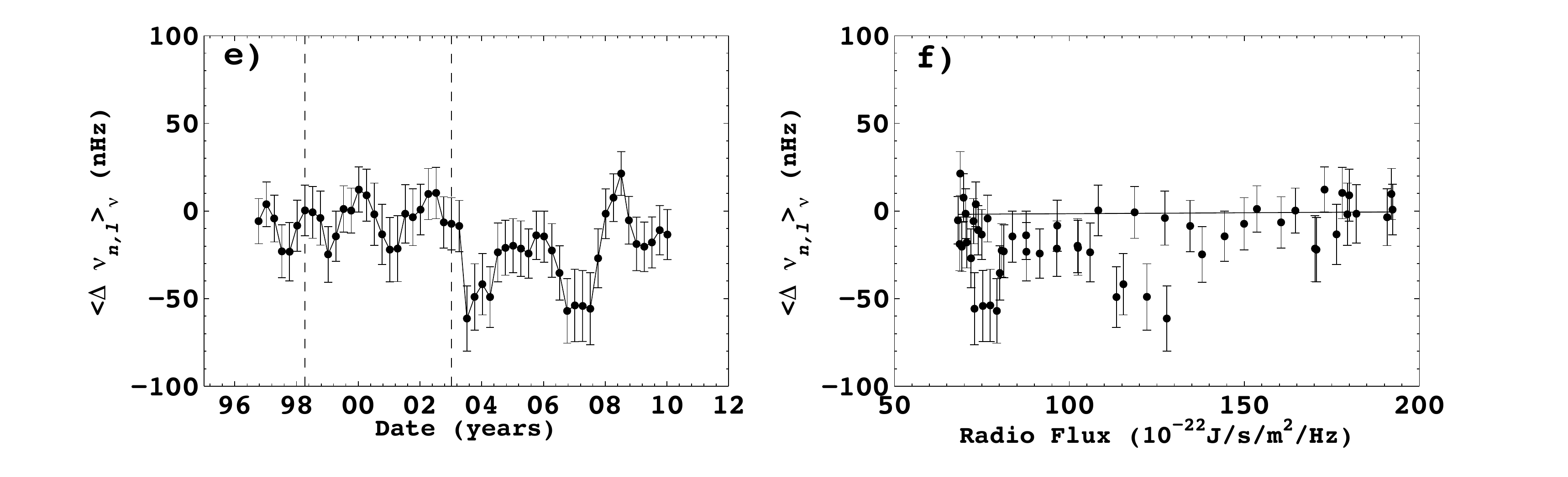}
\includegraphics[scale=0.265]{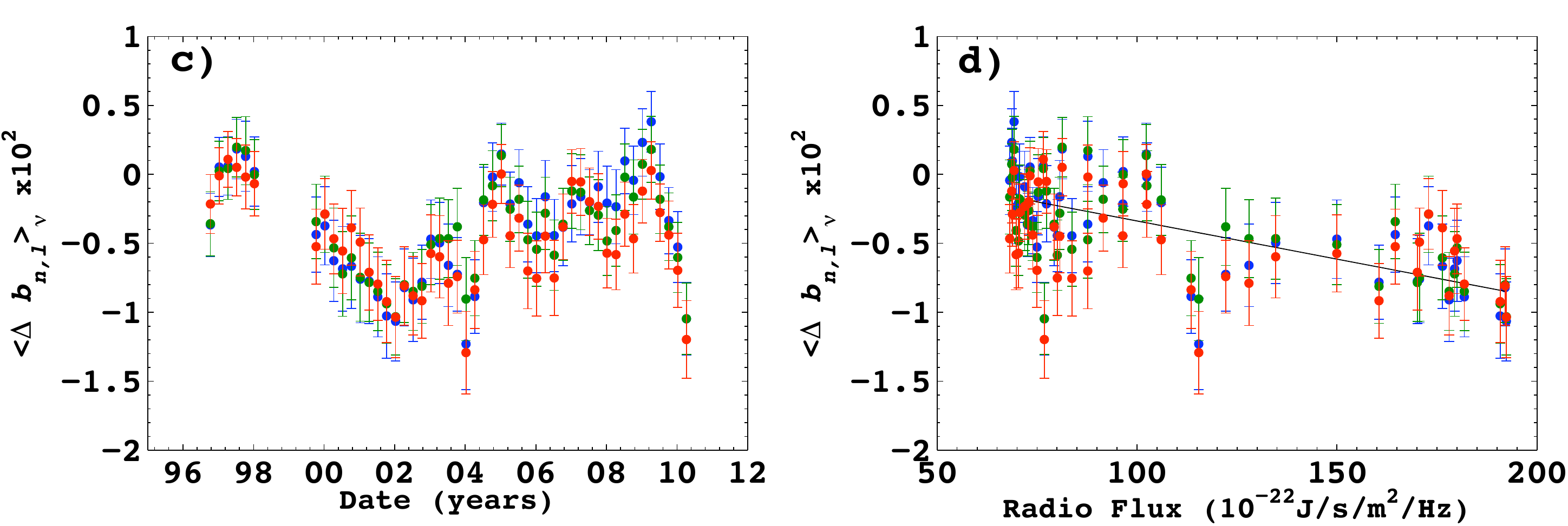}  \includegraphics[scale=0.265]{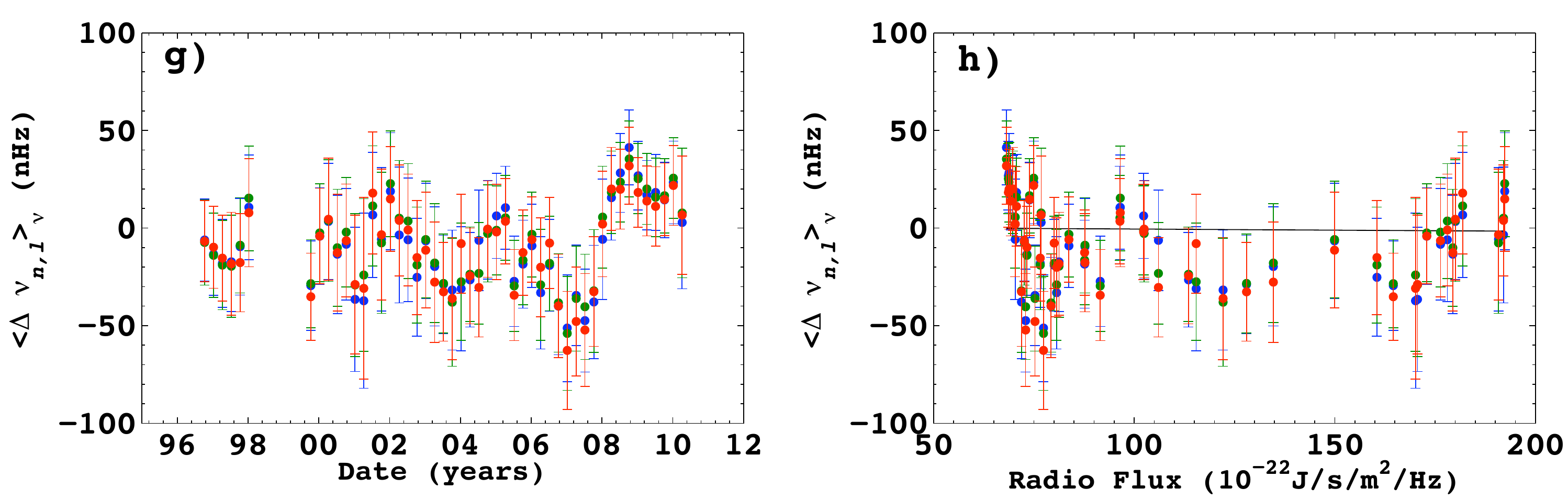} 
\end{center} 
\caption{\label{fig:golf_asym} (Top row) Temporal variations ($\times 10^2$) in mode asymmetry $\langle\Delta b\rangle$ (a, b) and in rotational splitting (nHz) (e, f) measured from GOLF data as a function of time and radio flux. The vertical dashed lines separate the blue-, red-, and blue-wing GOLF periods. (Bottom row) Temporal variations ($\times 10^2$) in mode asymmetry $\langle\Delta b\rangle$ (c, d) and in rotational splitting (nHz) (g, h) measured from VIRGO data as a function of time and radio flux. The solid lines in panels b, d, f, and h represent the best fits from weighted linear regressions.} 
\end{figure}

\subsection{Frequencies of individual low-degree modes}
Different temporal variations of the Sun-as-a-star p-mode frequencies are observed between individual angular degrees $l = 0$, 1, and 2 (Salabert et al. [6]) and are consistent between the observations from GOLF and the 3 VIRGO/SPMs (Fig.~\ref{fig:golf_freq}). They might be interpreted as different geometrical responses to the spatial distribution of the solar magnetic field beneath the surface of the Sun. Moreover, the peculiar behavior of the individual modes around the long and deep activity minimum of cycle~23 is of particular interest. 

\subsection{Peak asymmetry}
The peak asymmetry $\langle\Delta b\rangle$ of the pairs $l$ = 0-2 and $l$ = 1-3 modes shows significant temporal variations with solar activity (Fig.~\ref{fig:golf_asym} and Table~\ref{tab:changeasym}). Similar changes are obtained between velocity (GOLF) and intensity (VIRGO) measurements, while Jim\'enez-Reyes et al. [15] found opposite variations. No significant correlation between asymmetry and signal-to-noise ratio is observed.  

\subsection{Rotational splittings}
No correlation is observed between the temporal variations of the mean $l = 1$ and 2 rotational splittings and the 11-year solar cycle (Fig.~\ref{fig:golf_asym}). However, some similar fluctuations between GOLF and VIRGO are present. Note also that the GOLF splittings show a jump between the red-wing period and the second blue-wing period starting end of 2002.

\section{Conclusions}
We analyzed more than 14 years of radial velocity and intensity helioseismic Sun-as-a-star data collected by the space-based GOLF and VIRGO instruments respectively to study the temporal variations of the low-degree p-mode parameters (excitation, damping, frequency, peak asymmetry, and rotational splitting) with solar activity. The observed changes in excitation and damping parameters confirm previous results. The frequency shifts present differences between individual $l$ modes, for instance showing different minima for cycle~23. The peak asymmetry shows significant and similar variations between radial velocity and intensity measurements. The rotational splittings do not show correlations with the 11-year solar magnetic cycle.

\ack
The GOLF and VIRGO instruments onboard SoHO are a cooperative effort of many individuals, to whom we are indebted. SoHO is a project of international collaboration between ESA and NASA. The 10.7-cm radio flux was obtained from the National Geophysical Data Center. DS acknowledges the support of the grant PNAyA2007-62650 from the Spanish National Research Plan. This work has been supported by the CNES/GOLF grant at the SAp CEA-Saclay.

\end{document}